# Megatrend and Intervention Impact Analyser for Jobs: A European Big Data Hackathon Entry

Rain Öpik[1], Toomas Kirt[2], Innar Liiv[1]

[1] Tallinn University of Technology, Akadeemia 15A, 12618 Tallinn, Estonia
rain.opik@gmail.com
innar.liiv@ttu.ee
[2] Statistics Estonia, Tatari 51, 10134 Tallinn
toomas.kirt@stat.ee

This paper presents results of a European Big Data Hackathon entry to facilitate data analysis and visualization of patterns in provided and external datasets about European jobs and skills mismatch and labour market in general. The main contributions of this work are: the development of a method to represent the complex labour market internal structure from the perspective of occupations sharing skills; developing and presenting the prototype, together with an extended description of constructing a graph and related necessary data processing. Since labour market is not an isolated phenomenon and is constantly impacted with external trends and interventions, the presented tool is designed to enable adding extra layers of external information: what is the impact of a megatrend or an intervention to the labour market? Which parts of labour market of what country is most vulnerable to approaching megatrend or planned intervention? A case of analysing the labour market together with the megatrend of automation and computerization of jobs is presented. The source code of the prototype is released as open source for better repeat-ability.

## 1. Introduction

The European Big Data Hackathon is an event organised by the European Commission (Eurostat) which gathered teams from all over Europe to compete for the best data product combining official statistics and big data to support policy makers in one pressing policy question facing Europe. The policy question for this year hackathon was set: How would you support the design of policies for reducing mismatch between jobs and skills at regional level in the EU through the use of data? (see details at: https://ec.europa.eu/eurostat/cros/EU-BD-Hackathon_en)

This paper presents results developed and delivered at that event: a prototype to visualize the labour market structure (based on ESCO) as a complex interdependency network graph to demonstrate the demand and supply match and mismatch (based on EURES Curriculum Vitae and Job Vacancy datasets). Visualization is possible at aggregated level as well as at regional level, allowing to link additional external information.

The main contributions of this work are:

- the development of a method to represent the complex labour market internal structure from the perspective of occupations sharing skills;
- developing and presenting the prototype, together with an extended description of constructing a graph and related necessary data processing;
- a case of visualizing jobs and skills mismatch together with an external in-formation (the jobs susceptible to computerization) in order to understand the interplay and patterns in several datasets jointly.

The source code of the prototype is released as open source for better repeatability at https://github.com/rainopik/eubdhack-megatrend and the prototype is available online at https://rainopik.github.io/eubdhack-megatrend/.

The rest of this paper is organized as follows: second section presents the motivation of why such kind of tools to link external solutions and research results, is necessary. In third section, a list of datasets used to develop the tool and analyse the case is presented. In forth section, the method for constructing the graph is presented. Finally, the prototype is presented and some concluding remarks.

## 2. Motivation

Every new megatrend creates a need for new policy. Every successful policy starts an intervention. Neither problems nor solutions of the labour market are only internal. New approaches and tools are necessary to understand the complex phenomena of the labour market (e.g. mismatch of job vacancies and people wanting different jobs), far beyond the level of slogans of different impact of megatrends to labour market (technological change, future of professions, automation and computerization of jobs, robots, urbanization, refugee crises etc.).

Recent advances in artificial intelligence (LeCun, Bengio and Hinton 2015) and automation have raised fears it will have significant impact on job market (Mitchell and Brynjolfsson 2017). For example, it was found that across the OECD countries, on average 9 % of jobs are automatable (Arntz, Gregory and Zierahn 2016). On the other hand, it does not mean that the jobs are disappearing fully, rather they are transferred to other jobs and industries that need different skills. As Lerman and Schmidt (2005) have found around the appearance of the first personal computers in the mid-seventies and 1983, computer industry jobs in the United States grew almost 80 percent, while total U.S. manufacturing employment increased by only 4 percent. But recent developments in technology affect too many industries simultaneously and it may cause accumulation of problems like it was the case with the year 2000 problem. We propose a tool which can be used to show structural changes visually. It belongs to the class of the anomaly detection tools, which are used for finding patterns in data that do not comply with expected behaviour (Chandola, Banerjee and Kumar 2009). For this project, the computerization of jobs was selected as a case study, but the application of the tool is much broader.

Our tool is universal and allows adding extra layers of information: what is the impact of a megatrend or an intervention to the labour market? Which parts of labour market of what country is most vulnerable to approaching megatrend or planned intervention?

To complete the case study for the prototype we had to link job demand and vacancy dataset with additional information: The future of employment: how susceptible are jobs to computerization? (Benedikt and Osborne 2017). To demonstrate the visualization of this megatrend on labour market, we had to extract data from scientific articles, recode O*NET-SOC standard to ISCO in order to link with datasets provided by European Big Data Hackathon.

Our tool allows visualization to understand the impact of computerization to job market. What occupations are most susceptible to computerization? Is it potentially going to hit a labour market demand and skills mismatch or accelerate the unemployment?

Recently, the topic of computerization of jobs is gaining a lot of media (Chui Manyika and Miremadi 2017; see corresponding report Manyika et al. 2017; Lohr 2017; see corresponding report National Academies of Sciences, Engineering, and Medicine 2017) and government (Furman et al. 2016) attention. However, the comparison of different statements and conclusions is difficult. Recent panel came to a conclusion (Lohr 2017) that new tools are necessary to track technology's impact. Natonial statistics offices need tools to comment and respond to different publications, opinions or other thought papers. This paper presents a case to link occupation similarity, job vacancy dataset and an opinion from a recent research paper (Benedikt and Osborne 2017).

## 3. Datasets

Several job related datasets were made available to the teams in the hackathon. To develop our prototype, we used CV and job vacancy data from EURES portal and the ESCO, the multilingual classification of European Skills, Competences, Qualifications and Occupations, datasets.

### *3.1. Primary datasets*

The basis of this paper is a list of jobs susceptible for Automation / Computerization (Benedikt and Osborne 2017), which outlines 702 occupations, classified in SOC, with a probability of computerisation in the near future.

For measuring the impact of computerisation, we chose to use the EURES dataset (see http://eures.europa.eu/), which provides insight of which jobs are offered by employers and looked by jobseekers across Europe. The EURES dataset consist of two main tables, one on anonymised curriculum vitae (4.7 million lines) proposed by jobseekers and another on job vacancies (35 million lines) published by potential employers. The vacancy dataset was extracted from the EURES database on December 2, 2016. As the organizers of the hackathon did not want to use all the data the same job vacancies were aggregated. The CV dataset included monthly downloaded CV-s from period of March 2015 to November 2016 and contained data of 297 940 unique jobseekers. Records in the CV table are classified by ESCO occupation identifiers, but the vacancy table is classified by ISCO identifiers.

### *3.2. Classifiers*

Since each of the three main datasets used a different structure for organizing occupations, we present an explanation of each classifier used.

ISCO (International Standard Classification of Occupations) is maintained by International Labour Organization and has been first published in 1957, with latest revision in 2008. While jobs in ISCO are organized into clearly defined set of groups according to tasks and duties undertaken in the job (International Labour Organization, 2008), the classifier does not explicitly provide a list of those tasks and duties in a machine-readable format.

ESCO (European Skills/Competences, qualifications and Occupations) is a relatively new classification system, first published in 2013 (European Commission 2013). The ESCO system provides occupational profiles showing the relationships between occupations, skills, competences and qualifications. It contains 619 ISCO and 2950 ESCO occupations, with references to map an occupation in the ESCO system to a corresponding job in ISCO. Besides an organization of occupations, ESCO provides a hierarchy of skills and competences. This paper has greatly benefited from 65814 relationships in the ESCO system that connects skills to occupations. Since the ESCO system was provided in a RDF (Resource Description Framework) format, we have decided to convert it to SQL to avoid use of graph databases which complicate the data processing pipeline.

SOC (Standard Occupational Classification) is a system maintained by U.S. Bureau of Labor Statistics, with latest revision published in 2010 (U.S. Bureau of Labor Statistics 2010).

To connect all datasets, we needed to convert the U.S based SOC job classifier to international system. For that purpose, we used an occupation classifications crosswalk table, which maps a O*NET-SOC occupation to a job in ISCO (Wojciech, Autor and Acemoglu 2016). Fig. 1 illustrates how all datasets are connected.

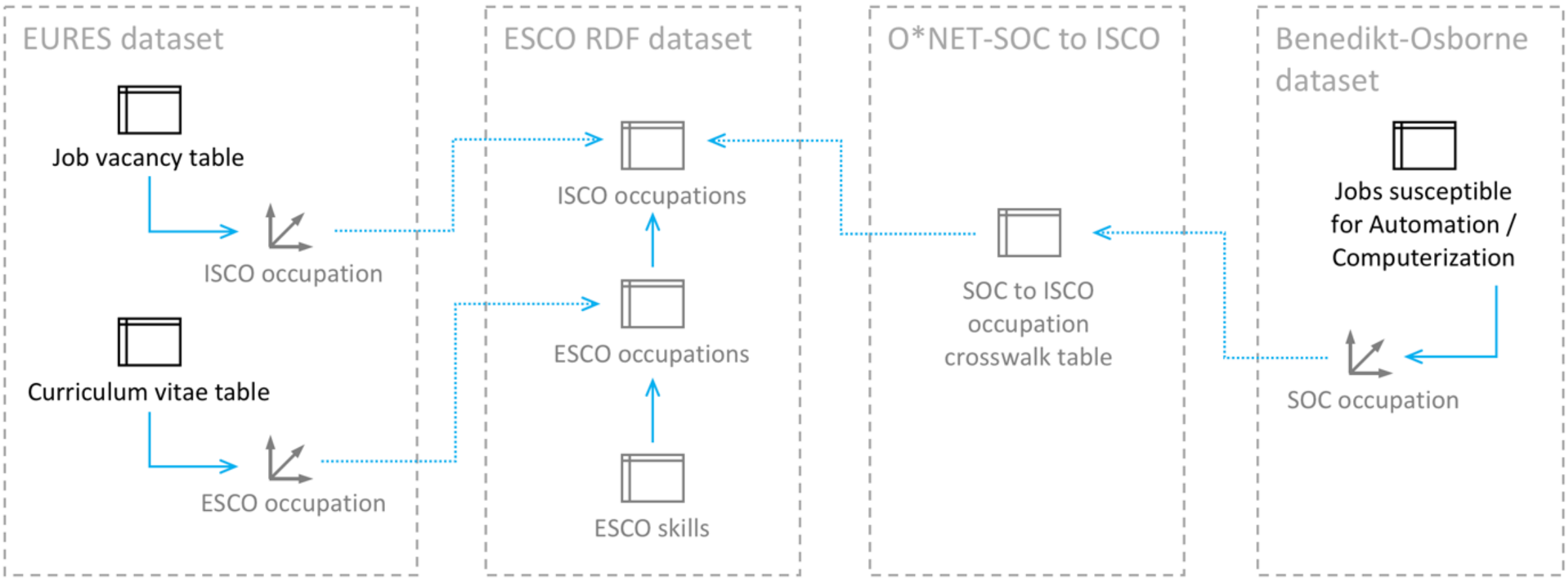

*Fig. 1. Overview of datasets and their relationships.*

## 4. Constructing a Graph

In order to visualize the complexity of a labour market, we propose to use graph theory (West 2001) to construct the node-link diagram (Collins, Penn and Carpendale 2009) to represent similarity and interrelations between different occupations in ESCO classification and dataset (European Commission 2013), according to shared skills required for that job. In next subsections we will present our approach and data modelling choices to construct the graph.

### *4.1. Occupation graph*

An occupation graph is defined by two entities: node and link. Node denotes an ESCO occupation (European Commission 2013). Each occupation may have additional data attributes attached to it. A link is defined between two nodes (occupations) when they are similar to each other. Therefore, the occupation graph is based similarity between ESCO occupations. We decided to define similarity based on skill information in the ESCO RDF classifier.

### *4.2. Linking similar occupations*

For a given ESCO occupation, we queried all skills that this occupation requires (relation type essential Skill in RDF). Then we matched all ESCO occupations that require the same skills. This produces a mapping ESCO occupation to ESCO occupation with a similarity measure that describes the ratio of shared skills between two occupations to number of all skills required by the first occupation.

Let us take two occupations: bus driver (ESCO occupation identifier: *00cee175-1376-43fb-9f02-ba3d7a910a58*) and private chauffeur (*e75305db-9011-4ee0-ab62-8d41a98f807e*) and enumerate all skills that are essential for both occupations.

*Table 1. A sample of essential skills for an occupation pair*

| Skills required for bus driver | Skills required for private chauffeur |
|---|---|
| provide first aid | N/A |
| manoeuvre bus | N/A |
| N/A | maintain personal hygiene standards |
| N/A | park vehicles |
| drive in urban areas | drive in urban areas |
| keep time accurately | keep time accurately |
| provide information to passengers | provide information to passengers |

The skills in this table can be divided into three groups:

- Skill that is only required for the first occupation (e.g. *bus driver*)

- Skill that is only required for the second occupation (*private chauffeur*)
- Skill that is required by both of these occupations.

When count the number of distinct skills that are required for both occupations (22 for this example) and divide it by the number of distinct skills required for the first occupation (35), we get a percentage of matching skills, which we can use as a similarity measure between these two occupations.

The resulting matrix is very large, as contains occupation pairs that are loosely connected by a very generic skill. For example, both *bus driver* and *physiotherapy assistant* have *use different communication channels* as essential skill, which connects them in the graph. However, when we calculate the skill match ratio, we get a modest 2%. Also the connection between these occupations does not make sense in real life, as it is difficult to imagine that a person skilled in operating heavy vehicles could easily apply for a position that requires medical skills.

To reduce the clutter, we decided to prune the graph of weakly connected occupation pairs and take only 3 most similar occupations for every occupation. Table 2 shows an example of the pruned graph for two selected occupations and Fig. 2 contains an illustration of how the graph would look like.

*Table 2. The pruned occupation graph for two occupations.*

| From occupation | To occupation | skill match |
|---|---|---:|
| bus driver | trolley bus driver | 80% |
| bus driver | tram driver | 77% |
| bus driver | private chauffeur | 63% |
| cargo vehicle driver | dangerous goods driver | 60% |
| cargo vehicle driver | bus driver | 55% |
| cargo vehicle driver | private chauffeur | 45% |

When this algorithm is run for all occupations (e.g. private chauffeur), we get new links in the graph, yielding at least three links for every node (Fig. 3).

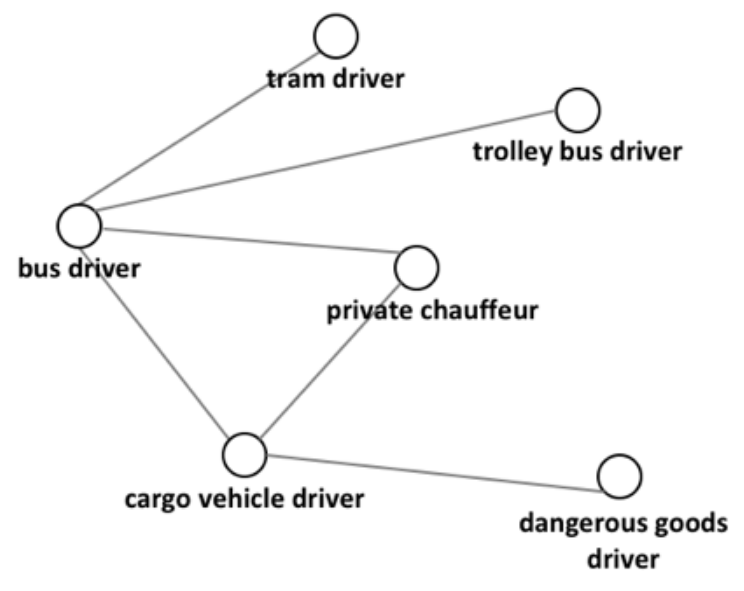

*Fig. 2. The graph for two occupations.*

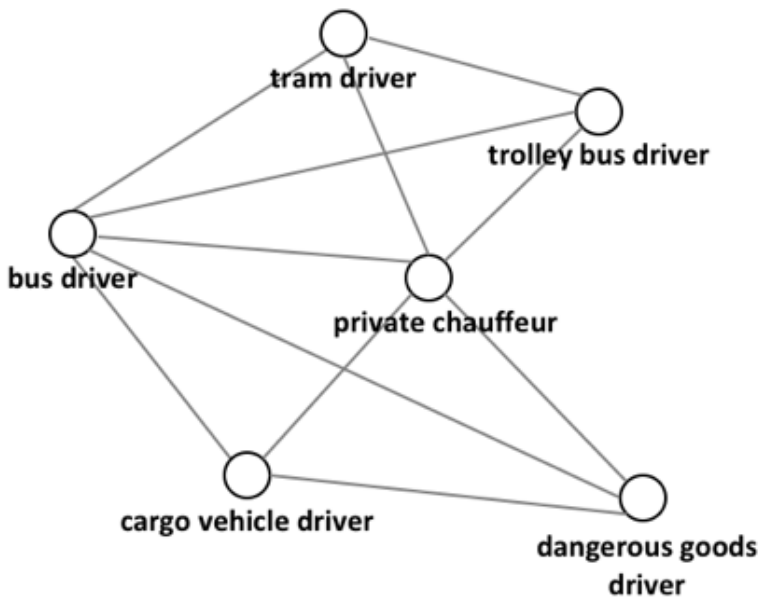

*Fig. 3. A fragment of the full occupation graph.*

### *4.3. Annotating occupations with supply and demand data*

Since each node in the occupation graph denotes ESCO occupation, we would like to know how this occupation will be affected by automation or computerization. The list of Jobs Susceptible for Automation (Benedikt and Osborne 2017) originally has SOC occupation codes. Mapping ISCO to SOC (Wojciech, Autor and Acemoglu 2016) is unfortunately one-to-many, which means that some ISCO occupations (e.g. *8332 - Heavy truck and lorry drivers*) are associated with several SOC occupations (*53-1031 - Driver/Sales Workers* and *53-3032 - Heavy and Tractor-Trailer Truck Drivers*) that may have differing probabilities for automation (respectively 0.98 and 0.79). To solve this ambiguity, we have calculated two probabilities, maximum and average.

After knowing, which jobs are going to be impacted, we wanted to assess, how many people would be affected by this trend. Since we have based our tool on EURES CV and job vacancy dataset, we could handily calculate the number of vacancies and number of unique persons that have marked this occupation as their desired job.

However, the amount of data in EURES dataset makes direct querying inefficient, therefore we decided to create a special crosstab table for storing occupation-based supply/demand counts by country. We used Apache Hive to run queries against EURES datasets and imported the results back to SQL.

For example, based on EURES data, there are 1925 job vacancies for bus driver in Austria and 5 job seekers have marked bus driver as their desired occupation - see Table 3 for example. Fig. 4 illustrates annotated nodes in a visual graph representation.

*Table 3. A node of data for two sample occupations.*

| Occupation | Prob. of automation | Vacancies total | CVs total | Vacancies in Austria | CVs in Austria | Vacancies in Belgium | CVs in Belgium | … |
|---|---|---|---|---|---|---|---|---|
| bus driver | 0.89 | 53 936 | 535 | 1 426 | | 1 925 | 5 | |
| cargo vehicle driver | 0.79 | 666 061 | 13 305 | 13 305 | | 35 475 | 15 | |
| … | | | | | | | | |

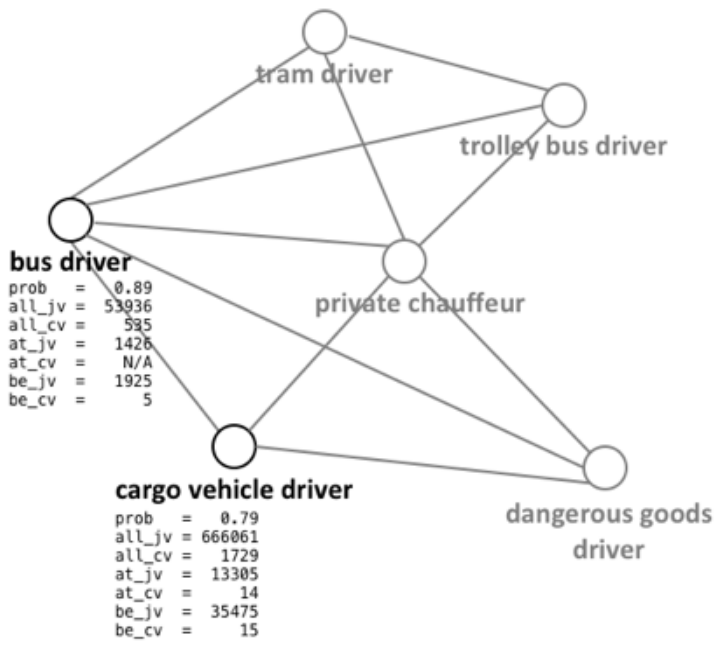

*Fig. 4. An occupation graph with annotations.*

## 5. Visualization Tool Prototype

### *5.1. Technical architecture*

Majority of the hackathon datasets were given as text files in a CSV format. After estimating the size of the main dataset (EURES), that was approximately 26 million records, we have decided to choose PostgreSQL for the primary database engine. In data exploration phase with relatively small number of data, relational databases have many benefits over specialized parallelized databases like Apache Hive, most important being the ease of ad-hoc querying and expressiveness of the SQL language.

The first draft of the occupation graph was drawn with python-graphtool (Peixoto 2014), which produced static image files. Since pre-rendered image files give a good overview of the graph, but lack on providing effective methods for filtering and obtaining details, we decided to implement the visualizer in d3.js (Bostock, Ogievetsky and Heer 2011). The d3.js application can be viewed in a modern web browser without any additional dependencies.

The visualizer tool was designed to run without server backend and online connection to database. This makes it easy to host the tool on a static website (like GitHub) without any running costs. The final table of similar occupations and list of all nodes in the graph were exported to text files so they can be served statically.

### *5.2. Data Processing*

We have built the occupation graph with PostgreSQL queries. The resulting graph was stored in two denormalized tables: *g_node* - containing a list of all occupations and their metadata and *g_link* – containing connections between similar occupations.

During construction of the node table, we have observed that counting the number of unique job seekers and vacancies by occupations and different countries is the most time-consuming part. Since this type of workload that is more suitable for database using the MapReduce programming model, we used Apache Hive to calculate the country-based aggregates for each occupation and gained tenfold increase in query performance.

The ESCO classifier was originally presented in a RDF format, which is a list of semantic triples in the subject-predicate-object format. While specialized graph databases have support querying data in the triple format (e.g. SPARQL or Gremlin), writing queries that join data across SQL and graph database is very inefficient performance wise. Therefore, we decided to parse the RDF file and convert it to relational structure suitable for SQL.

The serverless design of the visualizer mandates that the data files are accessible without a database. We have used flat CSV files for feeding data to the visualizer. Fig. 5 shows the data processing pipeline.

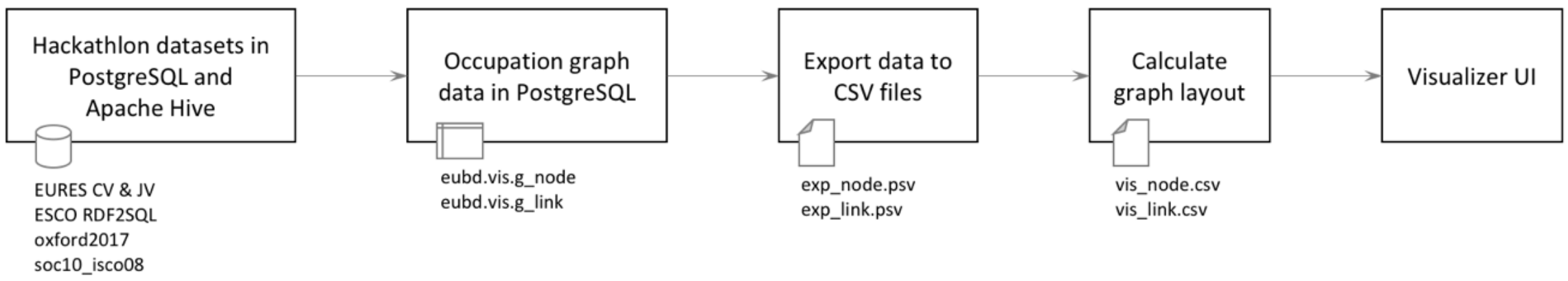

*Fig. 5. The data processing pipeline.*

### *5.3. Calculating graph layout*

Our experience with d3.js have shown that real-time calculation of graph layout (how to position nodes on a 2-dimensional plane) may be slow for graphs with non-trivial structure. Our occupation graph has 2950 nodes and 8838 links and after some experimentation we decided to pre-calculate the positions of graph nodes. We have used the SFDP layout algorithm (Yifan 2005) from graph-tool (Peixoto 2014) for calculating the position of nodes and re-indexing node identifiers to format that is suitable for visualizer.

Besides performance gains, this also has a second benefit – the visualization can be easily shared with fellow analysts. Most graph layout algorithms are non-deterministic in nature due to random initialization and produce a different layout after each run. By using pre-calculated node coordinates, we can make sure that visualizer produces output that looks exactly the same in every browser given the same set of input parameters.

### *5.4. Visualizer UI*

The user interface for visualizer is built with d3.js, which renders a zoomable and scrollable SVG document for browsing the graph online. The prototype application can be viewed in a modern web browser, preferably Google Chrome.

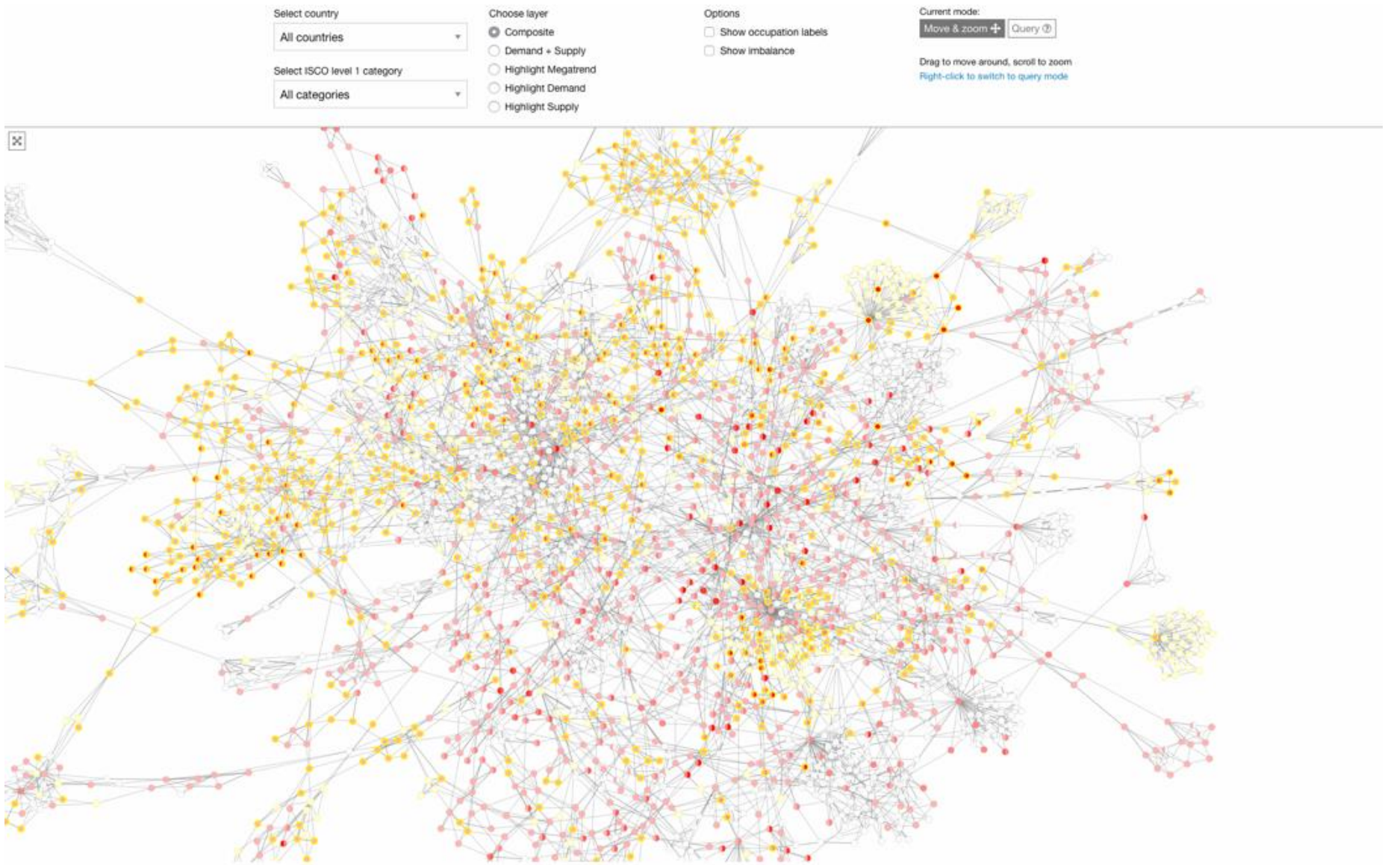

*Fig. 6. The visualizer prototype.*

## *5.5. Prototype interaction models*

Initially the visualizer displays the complete occupation graph. To facilitate getting more detailed information, the application has two modes:

- Move & zoom mode – analyst can click and drag mouse to move around the graph. Scrolling the mouse wheel zooms in and out.
- Query mode – when analyst moves the mouse cursor over a node, a small tooltip with demand and supply numbers will be displayed. Hovering also highlights connected jobs and fades out the rest of the graph. A click on the right mouse button allows switching between Move and Query modes. See Fig. 7 for example of query mode activated for the bus driver node.

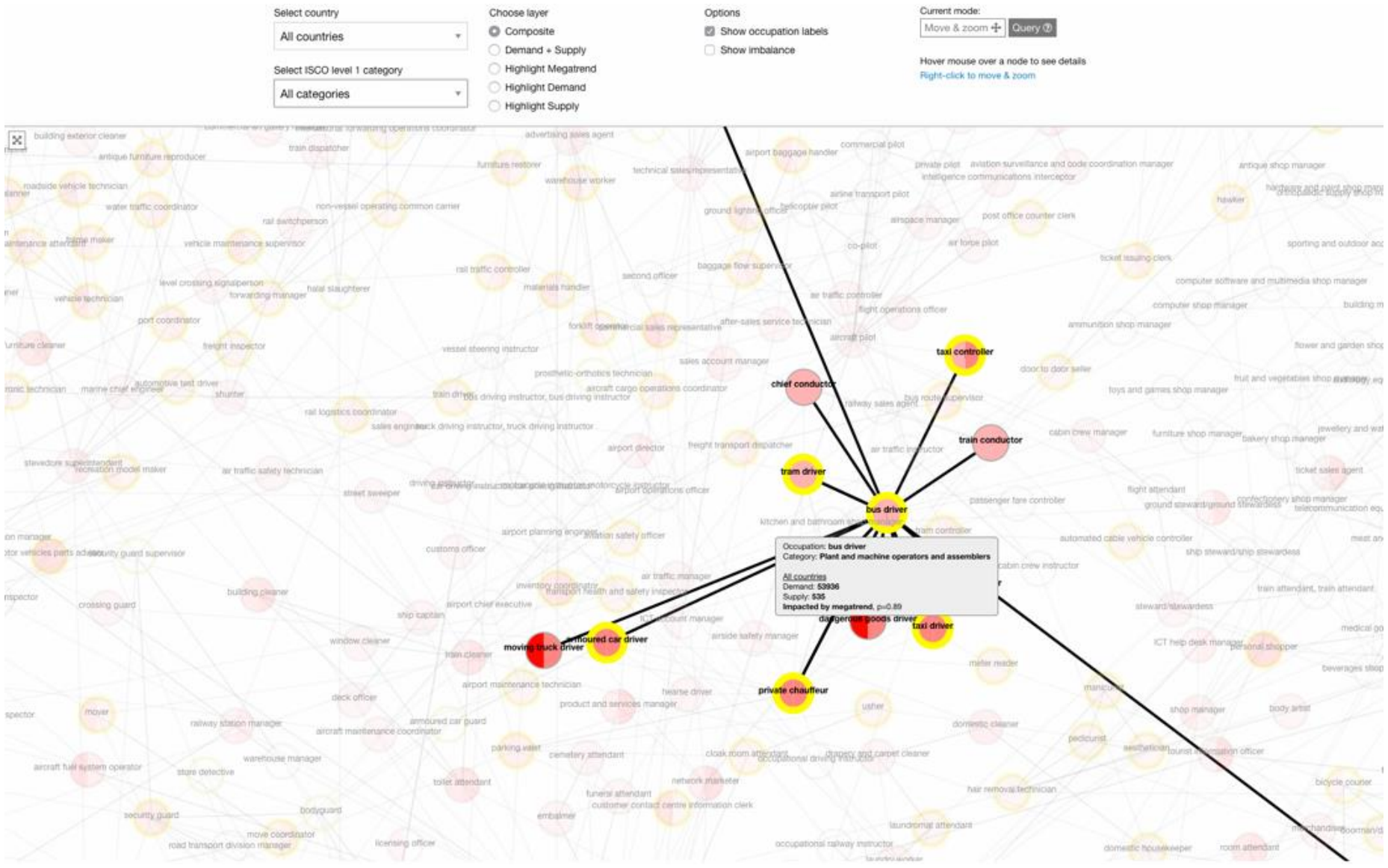

*Fig. 7. A query mode is activated for a node*

The full occupation graph has enough nodes to look like an impenetrable hairball when zoomed in. To reduce the clutter, we have added a filter tool to show only a relevant subset of occupations. Filter tool allows analyst to choose ISCO level 1 occupation category (e.g. *Plant and machine operators and assemblers*) and render only these occupations that have this categorization while hiding the rest of the graph. See Fig. 8 and Fig. 9 for the effect of the filter tool.

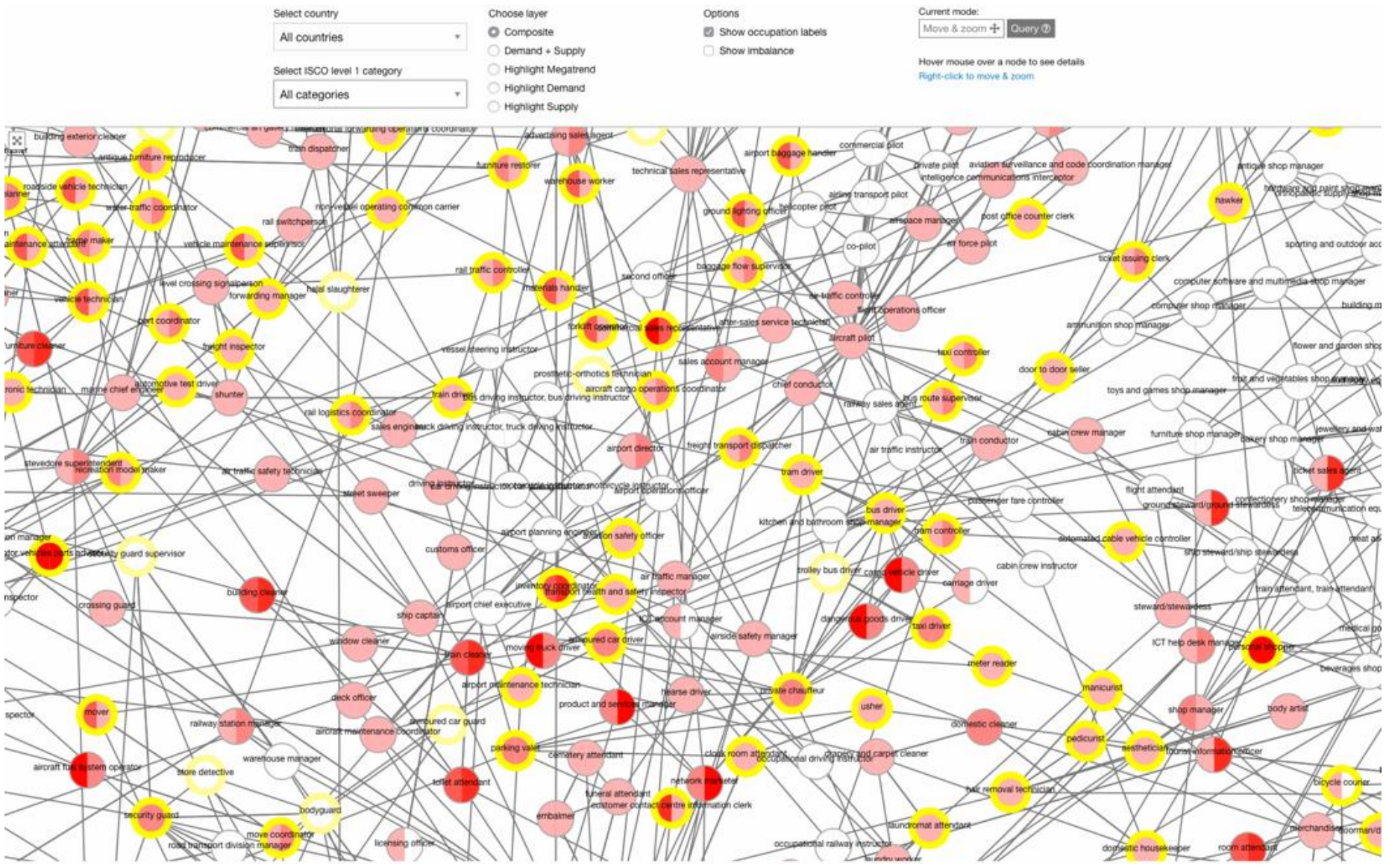

*Fig. 8. A close-up of the occupation graph.*

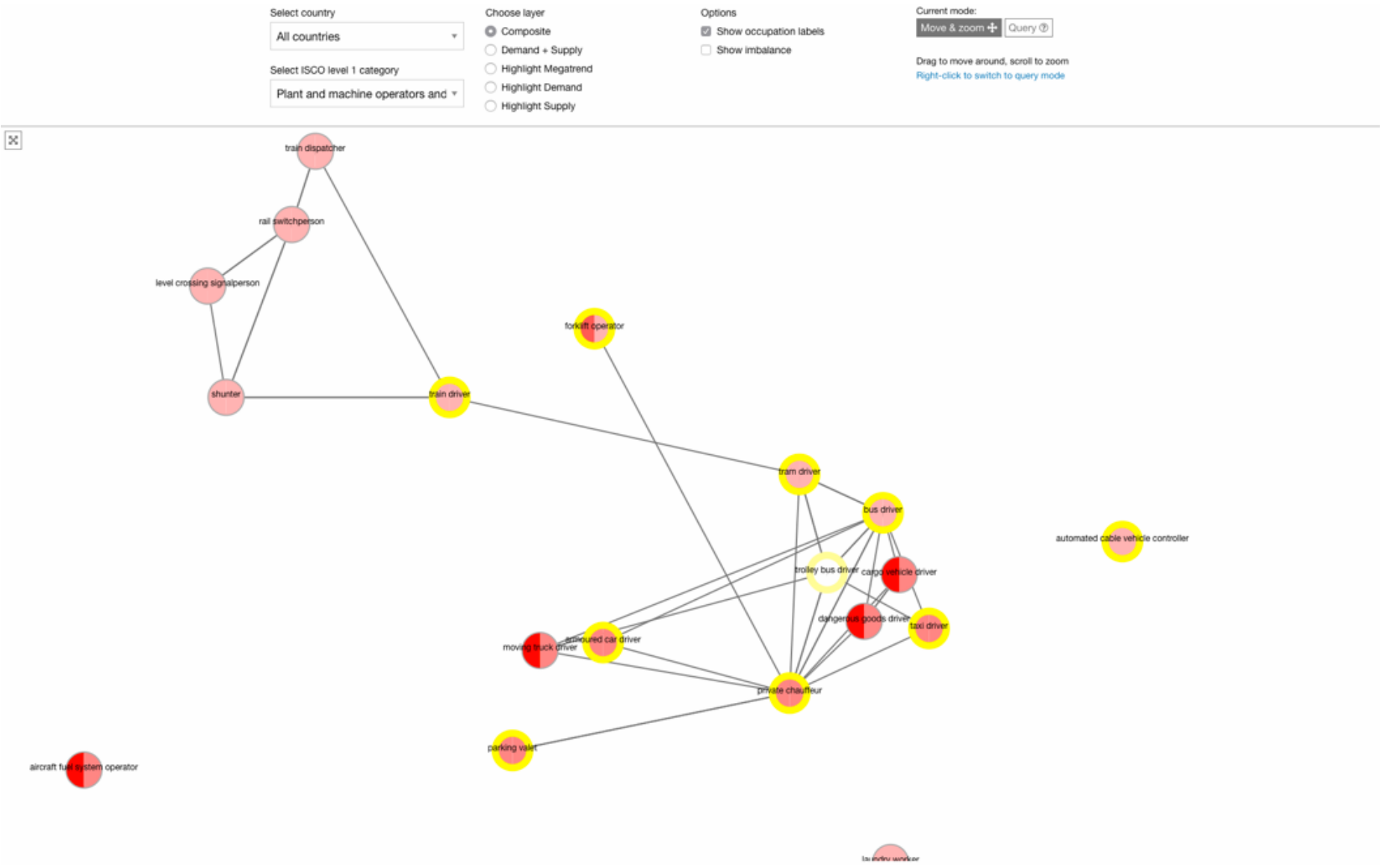

*Fig. 9. The filtering tool is applied.*

The filter tool is effective due to nature of the dataset – since nodes represent ESCO occupations which can be linked to hierarchical ISCO classification, the top level of the ISCO classifier produces a meaningful subset of the graph with same semantics.

## *5.6. Visualizing node metadata*

We have used colour to encode various metadata attributes that were attached to graph nodes. The visualizer supports several types for colour coding – we call them layers.

- Composite layer. Left half of a node is coloured by the number of vacancies available for that job (demand). White means no vacancies, light pink low and red denoting high demand. Right half is coloured by number of job seekers who have listed this job in their desired job list. Colour gradation is similar to the left half. Node is marked with a yellow halo when this job is affected by the Megatrend, i.e. job is in the list of jobs susceptible for automation / computerization (Benedikt and Osborne 2017).
- Demand and Supply layer. This is basically the same visualization as Composite, except the Megatrend markers (yellow halo) are not drawn.

- Highlight Megatrend layer. Node is coloured red when the job is affected by the Megatrend. Non-affected jobs are coloured white.
- Highlight Supply layer. Node is coloured red when at least one job seeker has listed this job in their desired job list. White nodes denote jobs that no-one desires.
- Highlight Demand layer. Node is coloured red when this job is listed in at least one job vacancy. White nodes denote jobs with no demand.

### *5.7. Demand & supply imbalance*

Colour values for the left and right half (demand and supply) are normalized separately due to huge imbalance in EURES data. Some countries have no job seekers in EURES while showing lots of vacancies and vice versa.

To overcome this issue, we have implemented an alternative way for colouring nodes. The default mode (*Show imbalance* unchecked) calculates the saturation ("brightness" of the red colour) of the left and the right half of the node on the same scale. This helps to identify the most demanded jobs – the analyst needs to look for nodes with a bright red left half. Similarly, jobs with the largest supply of seekers have a bright right half. For example, on Fig. 10, the occupation *Private chauffeur* has total demand (across all EU countries) 215677 and total supply of 1674. When these counts are normalized across the whole graph, both numbers are assigned a same colour.

Enabling the *Show imbalance* mode normalizes both colours on the same scale. This visualizes imbalance - when the left half of the node is brighter red compared to the right, this job has unsatisfied demand. Conversely, a brighter right half marks jobs with excessive of job seekers. Fig. 11 illustrates this.

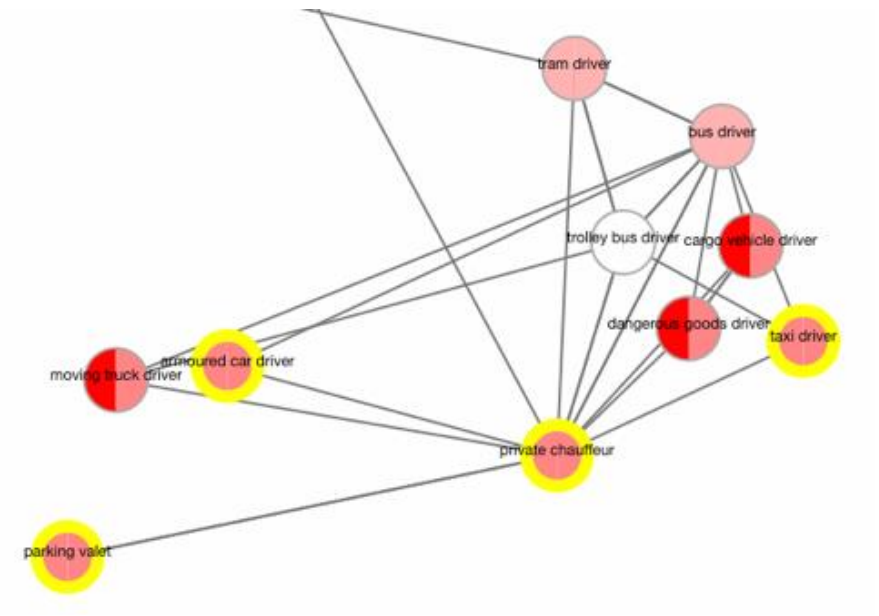

*Fig. 10. Default colouring of nodes*

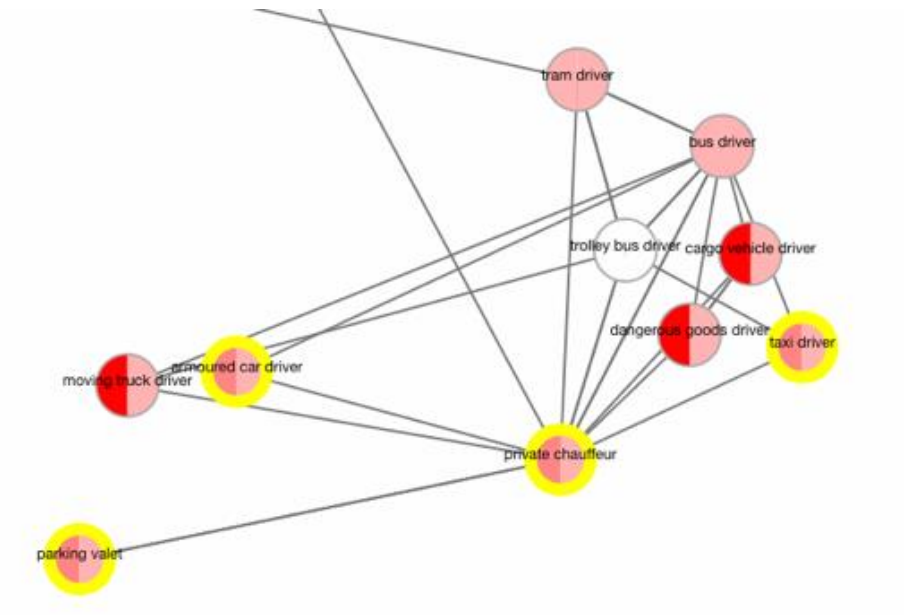

*Fig. 11. Show imbalance mode activated*

Note: the EURES data contains huge discrepancies between supply and demand across different countries. Some countries have no job seekers in EURES while showing lots of vacancies and vice versa. Therefore, the *Show imbalance* mode may reveal only the extremities.

## 6. Conclusion

In this paper, we presented the results of a European Big Data Hackathon Entry Megatrend and Intervention Impact Analyser for Jobs. The main contributions of this work were the development of a method to represent labour market structure and clusters from the perspective of occupations sharing skills, and presenting the proto-type, together with extended description of necessary data processing.

Rapid changes in a society in information age cause also challenges to national statistics offices. Registering time series might not be enough to face those challenges there might be needed some other approaches which indicate to some changes in the future and use in this purpose new sources of data. There is also need to educate people to read and understand the data visualized and represented in a new way. It is the direction the NSI-s are looking. There are first attempts to use big data sources and generate new type of statistics (e.g., traffic statistics based on the readings of traffic sensors (Daas et al. 2015)). With our tool we provide a way to foresee the changes in labour market. It is a first prototype and there are ways to improve it, as we could use longer time series and to follow and to visualize better the dynamics of skills. The job statistics by the field of occupations and skills taught in schools could also be used as additional data source. To identify quick changes in society we need new tools and need to use new data sources which help to react accurately and timely to them.